\documentclass[reprint,apl]{revtex4-1}
\usepackage{amssymb,amsmath}
\usepackage{graphicx}
\usepackage{dcolumn}
\usepackage{multirow}

\newcommand{\units}[1]{\ensuremath{\mathrm{#1}}}

\newcommand{\amount}[2]{\ensuremath{#1\:\units{#2}}}
\newcommand{\sym}[2]{\ensuremath{#1_{\mathrm{#2}}}}

\newcommand{\gmb}{\ensuremath{g\sym{\mu}{B}}}

\begin{document}

\title{\textbf{\fontfamily{phv}\selectfont Tunable singlet-triplet splitting in a few-electron Si/SiGe quantum dot}}
\author{Zhan Shi}
\author{C. B. Simmons}
\author{J. R. Prance}
\author{John King Gamble}
\author{Mark Friesen}
\author{D. E. Savage}
\author{M. G. Lagally}
\author{S. N. Coppersmith}
\author{M. A. Eriksson}
\affiliation{University of Wisconsin-Madison, Madison, WI 53706}

\begin{abstract}
We measure the excited-state spectrum of a Si/SiGe quantum dot as a function of in-plane magnetic field, and we identify the spin of the lowest three eigenstates in an effective two-electron regime.  The singlet-triplet splitting is an essential parameter describing spin qubits, and we extract this splitting from the data. We find it to be tunable by lateral displacement of  the dot, which is realized by changing two gate voltages on opposite sides of the device.  We present calculations showing the data are consistent with a spectrum in which the first excited state of the dot is a valley-orbit state.
\end{abstract}

\maketitle

Silicon quantum dots are candidate hosts for semiconductor spin qubits, both because of long spin relaxation and coherence times for electrons in Si, and because of potential synergy with classical microelectronics.  Long spin relaxation times have been demonstrated in Si quantum dots and donors~\cite{Xiao:2010p096801,Simmons:2011p156804,Hayes:2009preprint,Morello:2010p687}, and measurements of ensembles of donor-bound spins by electron spin resonance have demonstrated \sym{T}{2} coherence times up to two seconds~\cite{Tyryshkin:2011preprint}.  One of the key properties of silicon quantum dot spin qubits is the ability to tune in real-time tunnel rates and couplings between neighboring dots by controlling electrostatic gate voltages~\cite{Zimmerman:2007p033507,Simmons:2009p3234,Tracy:2010p192110}.  Tunable, gate-defined silicon quantum dots often are designed to sit at the interface between pure Si and a barrier of either SiGe~\cite{Slinker:2005p246,Berer:2006p162112} or SiO$_2$~\cite{Angus:2007p845,Nordberg:2009p202102,Xiao:2010p032103}.

For initialization and readout of singlet-triplet (S-T) spin qubits, an essential parameter is the energy difference \sym{E}{ST} between the singlet and triplet states of two electrons in one dot~\cite{Levy:2002p1446,Petta:2005p2180}. The energy \sym{E}{ST} is equal to the lowest single-particle excited state energy, less a correction arising from electron-electron interactions.  In silicon nanostructures, which have states arising from the two low-lying valleys in the Si conduction band, the sharpness and quality of the interface between Si and the SiGe or SiO$_2$ barrier material play an important role in determining this energy~\cite{Friesen:2006p202106}.  Experiments have shown that quantum-confined structures can have reasonably large valley splitting, ranging from \amount{100}{\mu eV} to \amount{1-2}{meV}~\cite{Goswami:2007p41,Borselli:2011p123118,Lim:2011preprint}. The existence of large valley splittings in silicon quantum dots have led to large \sym{E}{ST} and the observation of Pauli spin blockade~\cite{Shaji:2008p540,Liu:2008p073310}.  However, systematic control of the valley splitting or, more directly, \sym{E}{ST} has not been demonstrated in a silicon quantum dot.

\begin{figure}[h]
\includegraphics[width=8cm]{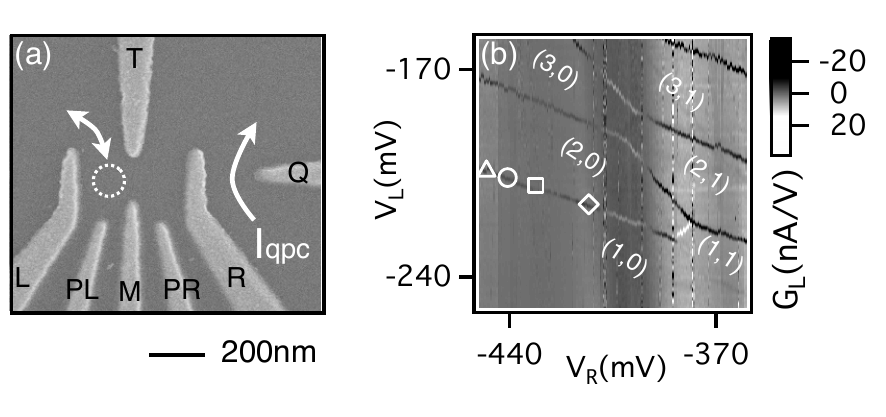}
\vspace{-12pt}
\caption{\label{fig1} (a) Scanning electron micrograph of a double dot identical to the one used in the experiment. The transition measured here is between the left dot (white circle) and the left reservoir. Charge sensing is performed by measuring the current $\sym{I}{qpc}$ through the QPC channel under a bias of \amount{500}{\mu V}.   (b) Stability diagram of the double dot with effective electron occupation numbers labeled.  The white symbols between regions (1,0) and (2,0) correspond to the gate voltages for the data reported below in Fig.~3.  The transition line at the bottom right  of the plot is invisible, because the tunnel rate between the right reservoir and the right dot is very slow in that gate voltage regime.}
\vspace{-12pt}
\end{figure}

In this letter, we report a magnetospectroscopy study of a Si/SiGe double quantum dot with 2 and 0 valence electrons on the left and right dots, respectively.  We use a pulsed-gate voltage technique to measure the evolution of the ground and low-lying excited states of the left dot as a function of an in-plane magnetic field $B$.  We extract the magnetic field \sym{B}{ST} at which the ground state changes from singlet to triplet, corresponding to the Zeeman energy equaling the singlet-triplet splitting \sym{E}{ST} for the $(2,0)$ charge configuration.  We find that \sym{B}{ST} is tunable by lateral displacement of the quantum dot location, achieved by simultaneously tuning voltages applied to two gates on opposite sides of the dot. \sym{B}{ST} evolves systematically as a function of the gate voltages, and we measure a fractional change in \sym{B}{ST} of up to \amount{19}{\%}. Changes in gate voltages can alter both the position and shape of the electron wavefunctions in quantum dots~\cite{Kyriakidis:2002p035320,Amasha:2008p2332,Thalakulam:2010p183104}.  Applying asymmetric changes to the voltages on either side of the quantum dot, as we do here, will change primarily the position of the quantum dot.  We perform calculations showing that the fractional change in \sym{B}{ST} we observe is consistent with valley-orbit mixing arising from a rough Si/SiGe interface, and that a change in position alone is sufficient to account for the magnitude of the observed changes in \sym{B}{ST}.

A double quantum dot, shown in Fig.~\ref{fig1}(a), is fabricated as described in~\cite{Simmons:2011p156804}. A quantum point contact is defined by gates R and Q and is used to perform charge sensing measurements. Gate L is connected to a pulse generator (Tektronix AFG 3252B), allowing the application of fast voltage pulses. The dc gate voltages are tuned so that the double dot is in the few-electron regime, as shown in Fig.~\ref{fig1}(b).  The change in background grayscale arises from changes in the QPC sensitivity caused by capacitive cross-talk in the device~\cite{Simmons:2007p213103}.  Measurements are performed in a dilution refrigerator at an electron temperature \amount{\sym{T}{e}=143 \pm 10}{mK}, determined as described in Ref.~\cite{Simmons:2011p156804}.  The electron occupation numbers are effective; we believe there are spin-zero closed shells of electrons in both the left and the right dots that do not participate in the physics discussed here.

\begin{figure}[h]
\includegraphics[width=8cm]{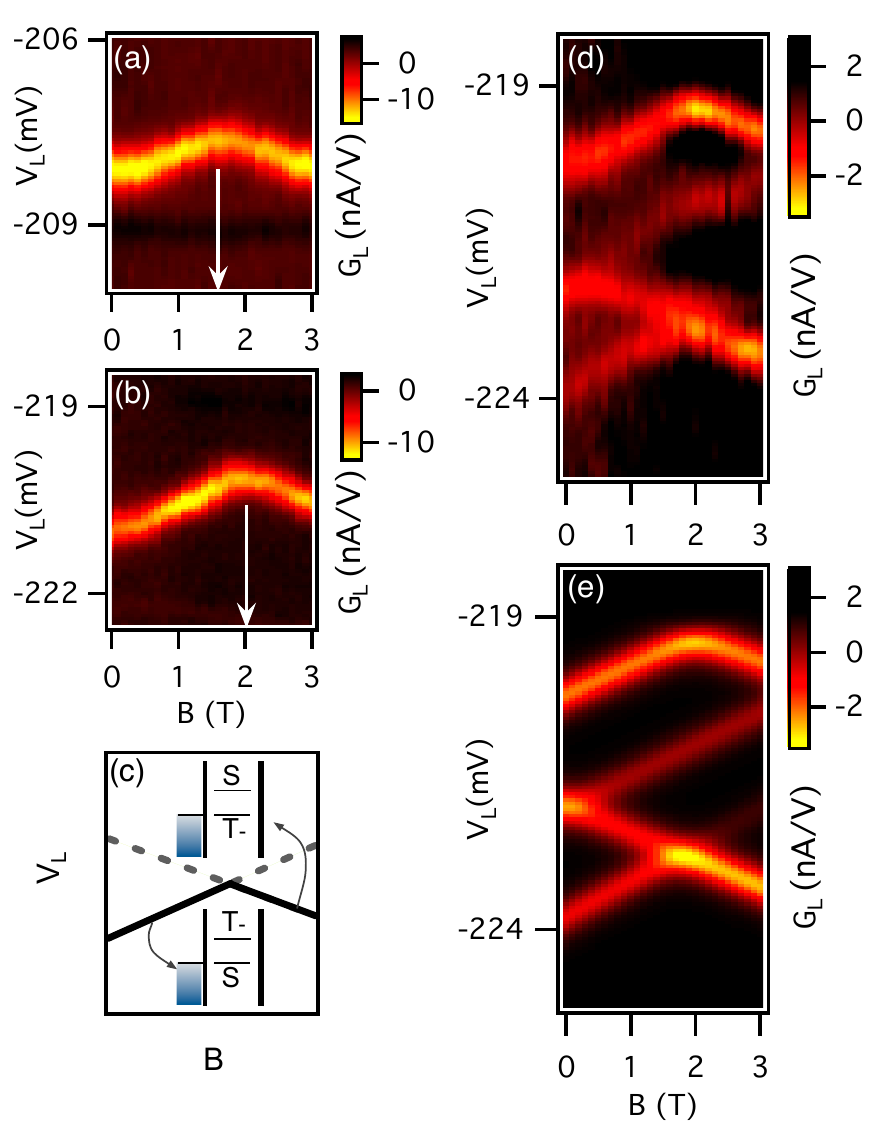}
\caption{\label{fig2} (a), (b) Ground state magnetospectroscopy for two different sets of gate voltages chosen so that the gate voltages for (b) favor a dot position farther to the right than those for (a) (see Fig.~\ref{fig3}). The positive slope for small $B$ corresponds to a singlet ground state, and the negative slope at large $B$ corresponds to a triplet T$_{-}$ ground state.  The turning point of the transition occurs when the Zeeman shift for the T$_{-}$ is equal to the zero-field \sym{E}{ST}. (c) Schematic diagram showing the transition as a function of $B$. (d) Excited state magnetospectroscopy using pulsed-gate voltages for the dot position corresponding to (b). Transitions to the three lowest states, S, T$_{-}$, and T$_{0}$ are clearly visible in the figure. (e) Simulated excited-state magnetospectroscopy for the data in panel (d).}
\vspace{-12pt}
\end{figure}

We determine the 2-electron singlet-triplet (S-T) splitting by using the charge sensing quantum point contact to measure the gate voltage dependence of the transition to the 2-electron state as a function of $B$. Figs.~\ref{fig2}(a) and (b) show the transconductance $\sym{G}{L}=\partial \sym{I}{qpc}/\partial\sym{V}{L}$ as a function of $B$, measured with a lock-in amplifier using a \amount{120}{\mu V} ac voltage applied to gate L.  The bright peak in the color plot corresponds to adding one electron to the left dot.  The gate voltage of this transition first increases and then decreases as a function of $B$.

The electrochemical potential $\sym{\mu}{N}$, and equivalently the gate voltage of transitions like those in Figs.~\ref{fig2}(a) and (b), has a dependence on the in-plane magnetic field of the form $\partial \sym{\mu}{N}/\partial B = \gmb \Delta \sym{S}{tot} (N)$~\cite{Hada:2003p155322}. Here $g$ is the Land\'{e} $g$-factor, $\sym{\mu}{B}$ is the Bohr magneton, and $\Delta\sym{S}{tot}(N)$ is the change in the $z$ component of the total spin when the $N$th electron is added to the dot. The electrochemical potential has a slope of $+\gmb/2$ when a spin-up electron is added (magnetic moment anti-parallel to $B$), whereas the addition of a spin-down electron results in a slope of $-\gmb/2$ (magnetic moment parallel to $B$).  The positive slope in Fig.~\ref{fig2}(a,b) at small $B$ corresponds to the addition of a spin-up electron, forming a 2-electron spin-singlet ground state.  At the value of $B$ marked with the arrows in panels (a) and (b), the slope changes; for $B$ larger than this value, the added electron is spin-down, and the ground state is the triplet T$_{-}$.  As indicated schematically in Fig.~\ref{fig2}(c), the turning point of the slope corresponds to a magnetic field $\sym{B}{ST}$ at which the Zeeman energy of the state T$_{-}$ is equal to $\sym{E}{ST}(B=0)$.  The value of \sym{B}{ST} is different in Figs.~\ref{fig2}(a) and (b), indicating that \sym{E}{ST} is tunable with gate voltage. We discuss the physics of this tunability in detail below.

Pulsed-gate spectroscopy methods~\cite{Elzerman:2004p731,Thalakulam:2011p045307} allow us to confirm the state identification described above, while simultaneously mapping out the excited-state energy spectrum as a function of $B$.  Square wave voltage pulses of peak-to-peak amplitude \amount{3.6}{mV} and frequency \amount{50}{kHz} are applied to gate L, and the time-averaged value of \sym{G}{L} is recorded as a function of $B$, as shown in Fig.~\ref{fig2}(d). 
Here, the bottom (top) line corresponds to the positive (negative) edge of the pulse bringing the 2-electron ground state into resonance with the Fermi level of the lead. Both of these lines therefore reproduce the shape of the line in Fig.~\ref{fig2}(b).

The two middle lines in Fig.~\ref{fig2}(d) meet at $B=0$ and correspond to the triplet states T$_{-}$ and T$_{0}$, which are degenerate at this point, and as $B$ increases, the lines split.  The T$_{-}$ line has a negative slope, corresponding to the addition of a spin-down electron, and this state becomes the ground state when $B=\sym{B}{ST}$.  The T$_{0}$ line has positive slope, corresponding to the addition of a spin-up electron.  The triplet T$_+$ state does not appear, as the addition of a single spin to the spin-down ground state cannot create a state with \amount{\sym{S}{Z}=+1}.

Figure 2(e) shows a theoretical simulation of the experiment of Fig.~\ref{fig2}(d), performed using a coupled rate equation model similar to that described in the supplemental material for Ref.~\cite{Simmons:2011p156804}.  The model includes thermal broadening but neglects energy-dependent tunneling. The S, T$_0$, and T$_-$ loading and unloading rates and the temperature are determined by fitting the simulation to the data in Fig.~\ref{fig2}(d).  We find the loading rates $\Gamma^L_S=45.1$ kHz, $\Gamma^L_{T_-}=216$ kHz, and $\Gamma^L_{T_0} = 377$ kHz, and the unloading rates $\Gamma^U_S=164$ kHz, $\Gamma^U_{T_-}=354$ kHz, and $\Gamma^U_{T_0} = 183$ kHz.

\begin{figure}[h]
\includegraphics[width=7cm]{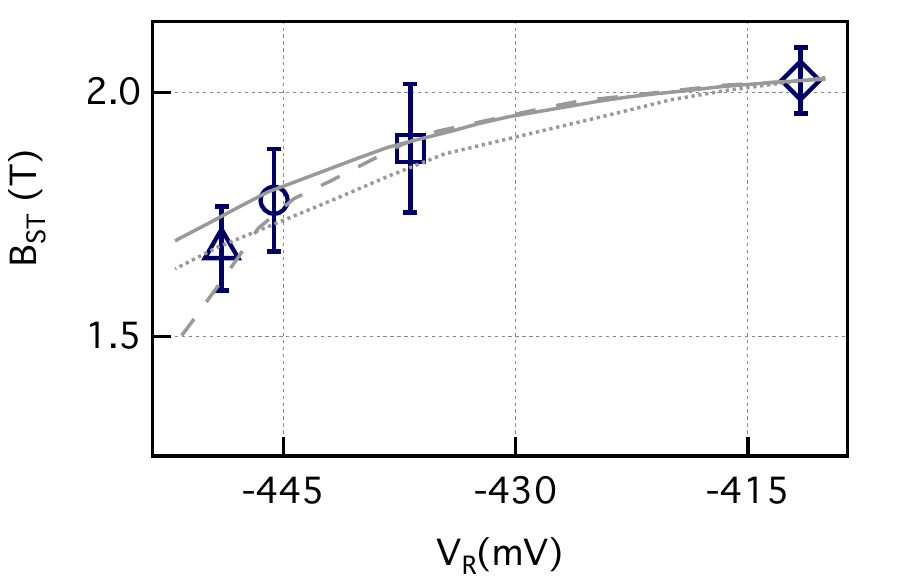}
\caption{\label{fig3} \sym{B}{ST}, the magnetic field at which the ground state shifts from singlet to triplet, for different sets of gate voltages $\sym{V}{L}$ and $\sym{V}{R}$, corresponding to the symbols on the stability diagram in Fig.~1(b).  Error bars are determined by the uncertainty in linear fits to lines like those in Fig.~2(a) and (b). The three gray lines correspond to three different sets of microscopic parameters that can be used to fit the experimental results.}
\vspace{-12pt}
\end{figure}

Using the method illustrated in Figs.~\ref{fig2}(a) and (b), we measure the transition field \sym{B}{ST} at four different gate voltage configurations, corresponding to the triangle, circle, square, and diamond shown in Fig.~\ref{fig1}(b).  Along this line in the stability diagram, changes in \sym{V}{L} and \sym{V}{R} tend to shift the dot physically from left to right as \sym{V}{R} (\sym{V}{L}) is made more positive (negative).  As shown in Fig.~\ref{fig3}, we observe a systematic increase in \sym{B}{ST} as we move from left to right in the stability diagram.  Over this range, \sym{B}{ST} increases from \amount{1.68\pm0.09}{T} to \amount{2.02\pm0.07}{T}, a total increase of about \amount{19}{\%}. 

The energy of singlet-triplet splitting can be expressed as $\sym{E}{ST} = \sym{E}{1}-\sym{E}{0}+\sym{C}{01}-\sym{C}{00}+\sym{K}{ST}$, where $\sym{E}{0}$ and $\sym{E}{1}$ denote the ground and first excited-state energies, \sym{C}{01} and \sym{C}{00} are the Coulomb interaction energies of the two electrons in the singlet and triplet states, and \sym{K}{ST} is the exchange energy~\cite{Hanson:2007p1217}.  A simple shift of the dot position is expected to have little effect on the last three terms, which correspond to interactions between the electrons.  Similarly, the shape of the wavefunction envelope should change very little as a function of the dot displacement.  Thus, it is important to check whether microscopic features of the quantum device can account for the changes in \sym{E}{ST} that we observe.

The single-particle spacing $\Delta E = \sym{E}{1} - \sym{E}{0}$ has a contribution arising from the difference in valley components in the two lowest lying orbital states.  The quantum well interface will have atomic steps and other sharp changes in potential that vary as a function of lateral position, and these variations can modify the coupling of the two z-valleys, contributing to a position dependence of the energy difference $\Delta E$~\cite{Friesen:2010p115324,Culcer:2010p205315}. 

To test whether a small atomic-scale variation can account for the magnitude of the observed variations in \sym{B}{ST}, we perform tight-binding calculations of the single particle energy levels of an electron confined near a single atomic step.  The calculations use a two-dimensional tight-binding Hamiltonian similar to Refs.~\cite{Boykin:2004p115,Saraiva:2010p245314}, including a parabolic lateral confinement potential.  The fitting procedure varies the position of the atomic step, the parabolic confinement length scale, and the vertical electric field, enabling a calculation of the variation in $\Delta E$ as a function of gate voltage.  To compare with the measured \sym{B}{ST}, we also fit the sum of the Coulomb and exchange energies ($\sym{C}{01}-\sym{C}{00}+\sym{K}{ST}$), and the results are plotted in Fig.~\ref{fig3}.  The fitting is underconstrained, as there are many physical ways to produce similar valley splitting.  To indicate the types of variations possible, three results are plotted in Fig.~\ref{fig3} as the solid, dashed, and dotted lines, and all three calculations can reproduce the magnitude of the observed change in \sym{B}{ST}. The the lowest excited-state can be classified as ``orbital-like'' when the calculated wavefunction contains a lateral node or
``valley-like'' when it does not, and both cases occur. For valley-like excitations, lateral translation of the dot with respect to a step results in a tunable valley splitting. For orbital-like excitations, strong valley-orbit coupling enables a tunable orbital energy splitting~\cite{Gamble:2011preprint}.

In conclusion, by performing magnetospectroscopy in a double quantum dot in Si/SiGe at the $(1,0)$ to $(2,0)$ charging transition, we measure the singlet-triplet splitting for the $(2,0)$ charge configuration and find it to be tunable by displacement of  the dot. Theoretical calculations show that atomic-scale structure of the quantum well interface is sufficient to produce valley-orbit mixing large enough to account for the experimental observations, even under the assumption that the dot shape is unchanged as a function of gate voltage.  The change in \sym{E}{ST} is large enough to be easily measured and offers an additional control knob for manipulation and initialization of singlet-triple quantum dot spin qubits.

We acknowledge experimental assistance from B. Rosemeyer and D. Greenheck.  This work was supported in part by ARO and LPS (W911NF-08-1-0482), NSF (DMR-0805045), and the United States Department of Defense. The views and conclusions contained in this document are those of the authors and should not be interpreted as representing the official policies, either expressly or implied, of the US Government.  This research utilized NSF-supported shared facilities at the University of Wisconsin-Madison.

%\bibliography{siliconqc,preprintsqc}

\begin{thebibliography}{34}%
\makeatletter
\providecommand \@ifxundefined [1]{%
 \@ifx{#1\undefined}
}%
\providecommand \@ifnum [1]{%
 \ifnum #1\expandafter \@firstoftwo
 \else \expandafter \@secondoftwo
 \fi
}%
\providecommand \@ifx [1]{%
 \ifx #1\expandafter \@firstoftwo
 \else \expandafter \@secondoftwo
 \fi
}%
\providecommand \natexlab [1]{#1}%
\providecommand \enquote  [1]{``#1''}%
\providecommand \bibnamefont  [1]{#1}%
\providecommand \bibfnamefont [1]{#1}%
\providecommand \citenamefont [1]{#1}%
\providecommand \href@noop [0]{\@secondoftwo}%
\providecommand \href [0]{\begingroup \@sanitize@url \@href}%
\providecommand \@href[1]{\@@startlink{#1}\@@href}%
\providecommand \@@href[1]{\endgroup#1\@@endlink}%
\providecommand \@sanitize@url [0]{\catcode `\\12\catcode `\$12\catcode
  `\&12\catcode `\#12\catcode `\^12\catcode `\_12\catcode `\%12\relax}%
\providecommand \@@startlink[1]{}%
\providecommand \@@endlink[0]{}%
\providecommand \url  [0]{\begingroup\@sanitize@url \@url }%
\providecommand \@url [1]{\endgroup\@href {#1}{\urlprefix }}%
\providecommand \urlprefix  [0]{URL }%
\providecommand \Eprint [0]{\href }%
\providecommand \doibase [0]{http://dx.doi.org/}%
\providecommand \selectlanguage [0]{\@gobble}%
\providecommand \bibinfo  [0]{\@secondoftwo}%
\providecommand \bibfield  [0]{\@secondoftwo}%
\providecommand \translation [1]{[#1]}%
\providecommand \BibitemOpen [0]{}%
\providecommand \bibitemStop [0]{}%
\providecommand \bibitemNoStop [0]{.\EOS\space}%
\providecommand \EOS [0]{\spacefactor3000\relax}%
\providecommand \BibitemShut  [1]{\csname bibitem#1\endcsname}%
\let\auto@bib@innerbib\@empty
%</preamble>
\bibitem [{\citenamefont {Xiao}\ \emph
  {et~al.}(2010{\natexlab{a}})\citenamefont {Xiao}, \citenamefont {House},\
  and\ \citenamefont {Jiang}}]{Xiao:2010p096801}%
  \BibitemOpen
  \bibfield  {author} {\bibinfo {author} {\bibfnamefont {M.}~\bibnamefont
  {Xiao}}, \bibinfo {author} {\bibfnamefont {M.~G.}\ \bibnamefont {House}}, \
  and\ \bibinfo {author} {\bibfnamefont {H.~W.}\ \bibnamefont {Jiang}},\ }\href
  {\doibase 10.1103/PhysRevLett.104.096801} {\bibfield  {journal} {\bibinfo
  {journal} {Phys. Rev. Lett.}\ }\textbf {\bibinfo {volume} {104}},\ \bibinfo
  {pages} {096801} (\bibinfo {year} {2010}{\natexlab{a}})}\BibitemShut
  {NoStop}%
\bibitem [{\citenamefont {Simmons}\ \emph {et~al.}(2011)\citenamefont
  {Simmons}, \citenamefont {Prance}, \citenamefont {Van~Bael}, \citenamefont
  {Koh}, \citenamefont {Shi}, \citenamefont {Savage}, \citenamefont {Lagally},
  \citenamefont {Joynt}, \citenamefont {Friesen}, \citenamefont {Coppersmith},\
  and\ \citenamefont {Eriksson}}]{Simmons:2011p156804}%
  \BibitemOpen
  \bibfield  {author} {\bibinfo {author} {\bibfnamefont {C.~B.}\ \bibnamefont
  {Simmons}}, \bibinfo {author} {\bibfnamefont {J.~R.}\ \bibnamefont {Prance}},
  \bibinfo {author} {\bibfnamefont {B.~J.}\ \bibnamefont {Van~Bael}}, \bibinfo
  {author} {\bibfnamefont {T.~S.}\ \bibnamefont {Koh}}, \bibinfo {author}
  {\bibfnamefont {Z.}~\bibnamefont {Shi}}, \bibinfo {author} {\bibfnamefont
  {D.~E.}\ \bibnamefont {Savage}}, \bibinfo {author} {\bibfnamefont {M.~G.}\
  \bibnamefont {Lagally}}, \bibinfo {author} {\bibfnamefont {R.}~\bibnamefont
  {Joynt}}, \bibinfo {author} {\bibfnamefont {M.}~\bibnamefont {Friesen}},
  \bibinfo {author} {\bibfnamefont {S.~N.}\ \bibnamefont {Coppersmith}}, \ and\
  \bibinfo {author} {\bibfnamefont {M.~A.}\ \bibnamefont {Eriksson}},\
  }\href@noop {} {\bibfield  {journal} {\bibinfo  {journal} {Phys. Rev. Lett.}\
  }\textbf {\bibinfo {volume} {106}},\ \bibinfo {pages} {156804} (\bibinfo
  {year} {2011})}\BibitemShut {NoStop}%
\bibitem [{\citenamefont {Hayes}\ \emph {et~al.}()\citenamefont {Hayes},
  \citenamefont {Kiselev}, \citenamefont {Borselli}, \citenamefont {Bui},
  \citenamefont {Croke}, \citenamefont {Deelman}, \citenamefont {Maune},
  \citenamefont {Milosavljevic}, \citenamefont {Moon}, \citenamefont {Ross},
  \citenamefont {Schmitz}, \citenamefont {Gyure},\ and\ \citenamefont
  {Hunter}}]{Hayes:2009preprint}%
  \BibitemOpen
  \bibfield  {author} {\bibinfo {author} {\bibfnamefont {R.~R.}\ \bibnamefont
  {Hayes}}, \bibinfo {author} {\bibfnamefont {A.~A.}\ \bibnamefont {Kiselev}},
  \bibinfo {author} {\bibfnamefont {M.~G.}\ \bibnamefont {Borselli}}, \bibinfo
  {author} {\bibfnamefont {S.~S.}\ \bibnamefont {Bui}}, \bibinfo {author}
  {\bibfnamefont {E.~T.}\ \bibnamefont {Croke}}, \bibinfo {author}
  {\bibfnamefont {P.~W.}\ \bibnamefont {Deelman}}, \bibinfo {author}
  {\bibfnamefont {B.~M.}\ \bibnamefont {Maune}}, \bibinfo {author}
  {\bibfnamefont {I.}~\bibnamefont {Milosavljevic}}, \bibinfo {author}
  {\bibfnamefont {J.-S.}\ \bibnamefont {Moon}}, \bibinfo {author}
  {\bibfnamefont {R.~S.}\ \bibnamefont {Ross}}, \bibinfo {author}
  {\bibfnamefont {A.~E.}\ \bibnamefont {Schmitz}}, \bibinfo {author}
  {\bibfnamefont {M.~F.}\ \bibnamefont {Gyure}}, \ and\ \bibinfo {author}
  {\bibfnamefont {A.~T.}\ \bibnamefont {Hunter}},\ }\href@noop {} {\enquote
  {\bibinfo {title} {Lifetime measurements ({T}$_1$) of electron spins in
  {S}i/{S}i{G}e quantum dots},}\ }\bibinfo {note} {ArXiv:0908.0173}\BibitemShut
  {NoStop}%
\bibitem [{\citenamefont {Morello}\ \emph {et~al.}(2010)\citenamefont
  {Morello}, \citenamefont {Pla}, \citenamefont {Zwanenburg}, \citenamefont
  {Chan}, \citenamefont {Tan}, \citenamefont {Huebl}, \citenamefont {Mottonen},
  \citenamefont {Nugroho}, \citenamefont {Yang}, \citenamefont {van Donkelaar},
  \citenamefont {Alves}, \citenamefont {Jamieson}, \citenamefont {Escott},
  \citenamefont {Hollenberg}, \citenamefont {Clark},\ and\ \citenamefont
  {Dzurak}}]{Morello:2010p687}%
  \BibitemOpen
  \bibfield  {author} {\bibinfo {author} {\bibfnamefont {A.}~\bibnamefont
  {Morello}}, \bibinfo {author} {\bibfnamefont {J.}~\bibnamefont {Pla}},
  \bibinfo {author} {\bibfnamefont {F.}~\bibnamefont {Zwanenburg}}, \bibinfo
  {author} {\bibfnamefont {K.}~\bibnamefont {Chan}}, \bibinfo {author}
  {\bibfnamefont {K.}~\bibnamefont {Tan}}, \bibinfo {author} {\bibfnamefont
  {H.}~\bibnamefont {Huebl}}, \bibinfo {author} {\bibfnamefont
  {M.}~\bibnamefont {Mottonen}}, \bibinfo {author} {\bibfnamefont
  {C.}~\bibnamefont {Nugroho}}, \bibinfo {author} {\bibfnamefont
  {C.}~\bibnamefont {Yang}}, \bibinfo {author} {\bibfnamefont {J.}~\bibnamefont
  {van Donkelaar}}, \bibinfo {author} {\bibfnamefont {A.}~\bibnamefont
  {Alves}}, \bibinfo {author} {\bibfnamefont {D.}~\bibnamefont {Jamieson}},
  \bibinfo {author} {\bibfnamefont {C.}~\bibnamefont {Escott}}, \bibinfo
  {author} {\bibfnamefont {L.}~\bibnamefont {Hollenberg}}, \bibinfo {author}
  {\bibfnamefont {R.}~\bibnamefont {Clark}}, \ and\ \bibinfo {author}
  {\bibfnamefont {A.}~\bibnamefont {Dzurak}},\ }\href@noop {} {\bibfield
  {journal} {\bibinfo  {journal} {Nature}\ }\textbf {\bibinfo {volume} {467}},\
  \bibinfo {pages} {687} (\bibinfo {year} {2010})}\BibitemShut {NoStop}%
\bibitem [{\citenamefont {Tyryshkin}\ \emph {et~al.}()\citenamefont
  {Tyryshkin}, \citenamefont {Tojo}, \citenamefont {Morton}, \citenamefont
  {Riemann}, \citenamefont {Abrosimov}, \citenamefont {Becker}, \citenamefont
  {Pohl}, \citenamefont {Schenkel}, \citenamefont {Thewalt}, \citenamefont
  {Itoh},\ and\ \citenamefont {Lyon}}]{Tyryshkin:2011preprint}%
  \BibitemOpen
  \bibfield  {author} {\bibinfo {author} {\bibfnamefont {A.~M.}\ \bibnamefont
  {Tyryshkin}}, \bibinfo {author} {\bibfnamefont {S.}~\bibnamefont {Tojo}},
  \bibinfo {author} {\bibfnamefont {J.~J.~L.}\ \bibnamefont {Morton}}, \bibinfo
  {author} {\bibfnamefont {H.}~\bibnamefont {Riemann}}, \bibinfo {author}
  {\bibfnamefont {N.~V.}\ \bibnamefont {Abrosimov}}, \bibinfo {author}
  {\bibfnamefont {P.}~\bibnamefont {Becker}}, \bibinfo {author} {\bibfnamefont
  {H.-J.}\ \bibnamefont {Pohl}}, \bibinfo {author} {\bibfnamefont
  {T.}~\bibnamefont {Schenkel}}, \bibinfo {author} {\bibfnamefont {M.~L.~W.}\
  \bibnamefont {Thewalt}}, \bibinfo {author} {\bibfnamefont {K.~M.}\
  \bibnamefont {Itoh}}, \ and\ \bibinfo {author} {\bibfnamefont {S.~A.}\
  \bibnamefont {Lyon}},\ }\href@noop {} {\enquote {\bibinfo {title} {Electron
  spin coherence exceeding seconds in high purity silicon},}\ }\bibinfo {note}
  {ArXiv:1105.3772v1}\BibitemShut {NoStop}%
\bibitem [{\citenamefont {Zimmerman}\ \emph {et~al.}(2007)\citenamefont
  {Zimmerman}, \citenamefont {Simonds}, \citenamefont {Fujiwara}, \citenamefont
  {Ono}, \citenamefont {Takahashi},\ and\ \citenamefont
  {Inokawa}}]{Zimmerman:2007p033507}%
  \BibitemOpen
  \bibfield  {author} {\bibinfo {author} {\bibfnamefont {N.~M.}\ \bibnamefont
  {Zimmerman}}, \bibinfo {author} {\bibfnamefont {B.~J.}\ \bibnamefont
  {Simonds}}, \bibinfo {author} {\bibfnamefont {A.}~\bibnamefont {Fujiwara}},
  \bibinfo {author} {\bibfnamefont {Y.}~\bibnamefont {Ono}}, \bibinfo {author}
  {\bibfnamefont {Y.}~\bibnamefont {Takahashi}}, \ and\ \bibinfo {author}
  {\bibfnamefont {H.}~\bibnamefont {Inokawa}},\ }\href {\doibase
  10.1063/1.2431778} {\bibfield  {journal} {\bibinfo  {journal} {Appl. Phys.
  Lett.}\ }\textbf {\bibinfo {volume} {90}},\ \bibinfo {pages} {033507}
  (\bibinfo {year} {2007})}\BibitemShut {NoStop}%
\bibitem [{\citenamefont {Simmons}\ \emph {et~al.}(2009)\citenamefont
  {Simmons}, \citenamefont {Thalakulam}, \citenamefont {Rosemeyer},
  \citenamefont {Bael}, \citenamefont {Sackmann}, \citenamefont {Savage},
  \citenamefont {Lagally}, \citenamefont {Joynt}, \citenamefont {Friesen},
  \citenamefont {Coppersmith},\ and\ \citenamefont
  {Eriksson}}]{Simmons:2009p3234}%
  \BibitemOpen
  \bibfield  {author} {\bibinfo {author} {\bibfnamefont {C.~B.}\ \bibnamefont
  {Simmons}}, \bibinfo {author} {\bibfnamefont {M.}~\bibnamefont {Thalakulam}},
  \bibinfo {author} {\bibfnamefont {B.~M.}\ \bibnamefont {Rosemeyer}}, \bibinfo
  {author} {\bibfnamefont {B.~J.~V.}\ \bibnamefont {Bael}}, \bibinfo {author}
  {\bibfnamefont {E.~K.}\ \bibnamefont {Sackmann}}, \bibinfo {author}
  {\bibfnamefont {D.~E.}\ \bibnamefont {Savage}}, \bibinfo {author}
  {\bibfnamefont {M.~G.}\ \bibnamefont {Lagally}}, \bibinfo {author}
  {\bibfnamefont {R.}~\bibnamefont {Joynt}}, \bibinfo {author} {\bibfnamefont
  {M.}~\bibnamefont {Friesen}}, \bibinfo {author} {\bibfnamefont {S.~N.}\
  \bibnamefont {Coppersmith}}, \ and\ \bibinfo {author} {\bibfnamefont {M.~A.}\
  \bibnamefont {Eriksson}},\ }\href@noop {} {\bibfield  {journal} {\bibinfo
  {journal} {Nano Lett.}\ }\textbf {\bibinfo {volume} {9}},\ \bibinfo {pages}
  {3234} (\bibinfo {year} {2009})}\BibitemShut {NoStop}%
\bibitem [{\citenamefont {Tracy}\ \emph {et~al.}(2010)\citenamefont {Tracy},
  \citenamefont {Nordberg}, \citenamefont {Young}, \citenamefont {Pinilla},
  \citenamefont {Stalford}, \citenamefont {Eyck}, \citenamefont {Eng},
  \citenamefont {Childs}, \citenamefont {Wendt}, \citenamefont {Grubbs},
  \citenamefont {Stevens}, \citenamefont {Lilly}, \citenamefont {Eriksson},\
  and\ \citenamefont {Carroll}}]{Tracy:2010p192110}%
  \BibitemOpen
  \bibfield  {author} {\bibinfo {author} {\bibfnamefont {L.~A.}\ \bibnamefont
  {Tracy}}, \bibinfo {author} {\bibfnamefont {E.~P.}\ \bibnamefont {Nordberg}},
  \bibinfo {author} {\bibfnamefont {R.~W.}\ \bibnamefont {Young}}, \bibinfo
  {author} {\bibfnamefont {C.~B.}\ \bibnamefont {Pinilla}}, \bibinfo {author}
  {\bibfnamefont {H.~L.}\ \bibnamefont {Stalford}}, \bibinfo {author}
  {\bibfnamefont {G.~A.~T.}\ \bibnamefont {Eyck}}, \bibinfo {author}
  {\bibfnamefont {K.}~\bibnamefont {Eng}}, \bibinfo {author} {\bibfnamefont
  {K.~D.}\ \bibnamefont {Childs}}, \bibinfo {author} {\bibfnamefont {J.~R.}\
  \bibnamefont {Wendt}}, \bibinfo {author} {\bibfnamefont {R.~K.}\ \bibnamefont
  {Grubbs}}, \bibinfo {author} {\bibfnamefont {J.}~\bibnamefont {Stevens}},
  \bibinfo {author} {\bibfnamefont {M.~P.}\ \bibnamefont {Lilly}}, \bibinfo
  {author} {\bibfnamefont {M.~A.}\ \bibnamefont {Eriksson}}, \ and\ \bibinfo
  {author} {\bibfnamefont {M.~S.}\ \bibnamefont {Carroll}},\ }\href@noop {}
  {\bibfield  {journal} {\bibinfo  {journal} {Appl. Phys. Lett.}\ }\textbf
  {\bibinfo {volume} {97}},\ \bibinfo {pages} {192110} (\bibinfo {year}
  {2010})}\BibitemShut {NoStop}%
\bibitem [{\citenamefont {Slinker}\ \emph {et~al.}(2005)\citenamefont
  {Slinker}, \citenamefont {Lewis}, \citenamefont {Haselby}, \citenamefont
  {Goswami}, \citenamefont {Klein}, \citenamefont {Chu}, \citenamefont
  {Coppersmith}, \citenamefont {Joynt}, \citenamefont {Blick}, \citenamefont
  {Friesen},\ and\ \citenamefont {Eriksson}}]{Slinker:2005p246}%
  \BibitemOpen
  \bibfield  {author} {\bibinfo {author} {\bibfnamefont {K.~A.}\ \bibnamefont
  {Slinker}}, \bibinfo {author} {\bibfnamefont {K.~L.~M.}\ \bibnamefont
  {Lewis}}, \bibinfo {author} {\bibfnamefont {C.~C.}\ \bibnamefont {Haselby}},
  \bibinfo {author} {\bibfnamefont {S.}~\bibnamefont {Goswami}}, \bibinfo
  {author} {\bibfnamefont {L.~J.}\ \bibnamefont {Klein}}, \bibinfo {author}
  {\bibfnamefont {J.~O.}\ \bibnamefont {Chu}}, \bibinfo {author} {\bibfnamefont
  {S.~N.}\ \bibnamefont {Coppersmith}}, \bibinfo {author} {\bibfnamefont
  {R.}~\bibnamefont {Joynt}}, \bibinfo {author} {\bibfnamefont {R.~H.}\
  \bibnamefont {Blick}}, \bibinfo {author} {\bibfnamefont {M.}~\bibnamefont
  {Friesen}}, \ and\ \bibinfo {author} {\bibfnamefont {M.~A.}\ \bibnamefont
  {Eriksson}},\ }\href {\doibase 10.1088/1367-2630/7/1/246} {\bibfield
  {journal} {\bibinfo  {journal} {New J. Phys.}\ }\textbf {\bibinfo {volume}
  {7}},\ \bibinfo {pages} {246} (\bibinfo {year} {2005})}\BibitemShut {NoStop}%
\bibitem [{\citenamefont {Berer}\ \emph {et~al.}(2006)\citenamefont {Berer},
  \citenamefont {Pachinger}, \citenamefont {Pillwein}, \citenamefont
  {M\"{u}hlberger}, \citenamefont {Lichtenberger}, \citenamefont {Brunthaler},\
  and\ \citenamefont {Sch\"{a}ffler}}]{Berer:2006p162112}%
  \BibitemOpen
  \bibfield  {author} {\bibinfo {author} {\bibfnamefont {T.}~\bibnamefont
  {Berer}}, \bibinfo {author} {\bibfnamefont {D.}~\bibnamefont {Pachinger}},
  \bibinfo {author} {\bibfnamefont {G.}~\bibnamefont {Pillwein}}, \bibinfo
  {author} {\bibfnamefont {M.}~\bibnamefont {M\"{u}hlberger}}, \bibinfo
  {author} {\bibfnamefont {H.}~\bibnamefont {Lichtenberger}}, \bibinfo {author}
  {\bibfnamefont {G.}~\bibnamefont {Brunthaler}}, \ and\ \bibinfo {author}
  {\bibfnamefont {F.}~\bibnamefont {Sch\"{a}ffler}},\ }\href {\doibase
  10.1063/1.2197320} {\bibfield  {journal} {\bibinfo  {journal} {Appl. Phys.
  Lett.}\ }\textbf {\bibinfo {volume} {88}},\ \bibinfo {pages} {162112}
  (\bibinfo {year} {2006})}\BibitemShut {NoStop}%
\bibitem [{\citenamefont {Angus}\ \emph {et~al.}(2007)\citenamefont {Angus},
  \citenamefont {Ferguson}, \citenamefont {Dzurak},\ and\ \citenamefont
  {Clark}}]{Angus:2007p845}%
  \BibitemOpen
  \bibfield  {author} {\bibinfo {author} {\bibfnamefont {S.~J.}\ \bibnamefont
  {Angus}}, \bibinfo {author} {\bibfnamefont {A.~J.}\ \bibnamefont {Ferguson}},
  \bibinfo {author} {\bibfnamefont {A.~S.}\ \bibnamefont {Dzurak}}, \ and\
  \bibinfo {author} {\bibfnamefont {R.~G.}\ \bibnamefont {Clark}},\ }\href
  {\doibase 10.1021/nl070949k} {\bibfield  {journal} {\bibinfo  {journal} {Nano
  Lett.}\ }\textbf {\bibinfo {volume} {7}},\ \bibinfo {pages} {2051} (\bibinfo
  {year} {2007})}\BibitemShut {NoStop}%
\bibitem [{\citenamefont {Nordberg}\ \emph {et~al.}(2009)\citenamefont
  {Nordberg}, \citenamefont {Stalford}, \citenamefont {Young}, \citenamefont
  {Eyck}, \citenamefont {Eng}, \citenamefont {Tracy}, \citenamefont {Childs},
  \citenamefont {Wendt}, \citenamefont {Grubbs}, \citenamefont {Stevens},
  \citenamefont {Lilly}, \citenamefont {Eriksson},\ and\ \citenamefont
  {Carroll}}]{Nordberg:2009p202102}%
  \BibitemOpen
  \bibfield  {author} {\bibinfo {author} {\bibfnamefont {E.~P.}\ \bibnamefont
  {Nordberg}}, \bibinfo {author} {\bibfnamefont {H.~L.}\ \bibnamefont
  {Stalford}}, \bibinfo {author} {\bibfnamefont {R.}~\bibnamefont {Young}},
  \bibinfo {author} {\bibfnamefont {G.~A.~T.}\ \bibnamefont {Eyck}}, \bibinfo
  {author} {\bibfnamefont {K.}~\bibnamefont {Eng}}, \bibinfo {author}
  {\bibfnamefont {L.~A.}\ \bibnamefont {Tracy}}, \bibinfo {author}
  {\bibfnamefont {K.~D.}\ \bibnamefont {Childs}}, \bibinfo {author}
  {\bibfnamefont {J.~R.}\ \bibnamefont {Wendt}}, \bibinfo {author}
  {\bibfnamefont {R.~K.}\ \bibnamefont {Grubbs}}, \bibinfo {author}
  {\bibfnamefont {J.}~\bibnamefont {Stevens}}, \bibinfo {author} {\bibfnamefont
  {M.~P.}\ \bibnamefont {Lilly}}, \bibinfo {author} {\bibfnamefont {M.~A.}\
  \bibnamefont {Eriksson}}, \ and\ \bibinfo {author} {\bibfnamefont {M.~S.}\
  \bibnamefont {Carroll}},\ }\href@noop {} {\bibfield  {journal} {\bibinfo
  {journal} {Appl. Phys. Lett.}\ }\textbf {\bibinfo {volume} {95}},\ \bibinfo
  {pages} {202102} (\bibinfo {year} {2009})}\BibitemShut {NoStop}%
\bibitem [{\citenamefont {Xiao}\ \emph
  {et~al.}(2010{\natexlab{b}})\citenamefont {Xiao}, \citenamefont {House},\
  and\ \citenamefont {Jiang}}]{Xiao:2010p032103}%
  \BibitemOpen
  \bibfield  {author} {\bibinfo {author} {\bibfnamefont {M.}~\bibnamefont
  {Xiao}}, \bibinfo {author} {\bibfnamefont {M.~G.}\ \bibnamefont {House}}, \
  and\ \bibinfo {author} {\bibfnamefont {H.~W.}\ \bibnamefont {Jiang}},\
  }\href@noop {} {\bibfield  {journal} {\bibinfo  {journal} {Appl. Phys.
  Lett.}\ }\textbf {\bibinfo {volume} {97}},\ \bibinfo {pages} {032103}
  (\bibinfo {year} {2010}{\natexlab{b}})}\BibitemShut {NoStop}%
\bibitem [{\citenamefont {Levy}(2002)}]{Levy:2002p1446}%
  \BibitemOpen
  \bibfield  {author} {\bibinfo {author} {\bibfnamefont {J.}~\bibnamefont
  {Levy}},\ }\href {\doibase 10.1103/PhysRevLett.89.147902} {\bibfield
  {journal} {\bibinfo  {journal} {Phys. Rev. Lett.}\ }\textbf {\bibinfo
  {volume} {89}},\ \bibinfo {pages} {147902} (\bibinfo {year}
  {2002})}\BibitemShut {NoStop}%
\bibitem [{\citenamefont {Petta}\ \emph {et~al.}(2005)\citenamefont {Petta},
  \citenamefont {Johnson}, \citenamefont {Taylor}, \citenamefont {Laird},
  \citenamefont {Yacoby}, \citenamefont {Lukin}, \citenamefont {Marcus},
  \citenamefont {Hanson},\ and\ \citenamefont {Gossard}}]{Petta:2005p2180}%
  \BibitemOpen
  \bibfield  {author} {\bibinfo {author} {\bibfnamefont {J.~R.}\ \bibnamefont
  {Petta}}, \bibinfo {author} {\bibfnamefont {A.~C.}\ \bibnamefont {Johnson}},
  \bibinfo {author} {\bibfnamefont {J.~M.}\ \bibnamefont {Taylor}}, \bibinfo
  {author} {\bibfnamefont {E.~A.}\ \bibnamefont {Laird}}, \bibinfo {author}
  {\bibfnamefont {A.}~\bibnamefont {Yacoby}}, \bibinfo {author} {\bibfnamefont
  {M.~D.}\ \bibnamefont {Lukin}}, \bibinfo {author} {\bibfnamefont {C.~M.}\
  \bibnamefont {Marcus}}, \bibinfo {author} {\bibfnamefont {M.~P.}\
  \bibnamefont {Hanson}}, \ and\ \bibinfo {author} {\bibfnamefont {A.~C.}\
  \bibnamefont {Gossard}},\ }\href {\doibase 10.1126/science.1116955}
  {\bibfield  {journal} {\bibinfo  {journal} {Science}\ }\textbf {\bibinfo
  {volume} {309}},\ \bibinfo {pages} {2180} (\bibinfo {year}
  {2005})}\BibitemShut {NoStop}%
\bibitem [{\citenamefont {Friesen}\ \emph {et~al.}(2006)\citenamefont
  {Friesen}, \citenamefont {Eriksson},\ and\ \citenamefont
  {Coppersmith}}]{Friesen:2006p202106}%
  \BibitemOpen
  \bibfield  {author} {\bibinfo {author} {\bibfnamefont {M.}~\bibnamefont
  {Friesen}}, \bibinfo {author} {\bibfnamefont {M.~A.}\ \bibnamefont
  {Eriksson}}, \ and\ \bibinfo {author} {\bibfnamefont {S.~N.}\ \bibnamefont
  {Coppersmith}},\ }\href {\doibase 10.1063/1.2387975} {\bibfield  {journal}
  {\bibinfo  {journal} {Appl. Phys. Lett.}\ }\textbf {\bibinfo {volume} {89}},\
  \bibinfo {pages} {202106} (\bibinfo {year} {2006})}\BibitemShut {NoStop}%
\bibitem [{\citenamefont {Goswami}\ \emph {et~al.}(2007)\citenamefont
  {Goswami}, \citenamefont {Slinker}, \citenamefont {Friesen}, \citenamefont
  {McGuire}, \citenamefont {Truitt}, \citenamefont {Tahan}, \citenamefont
  {Klein}, \citenamefont {Chu}, \citenamefont {Mooney}, \citenamefont {van~der
  Weide}, \citenamefont {Joynt}, \citenamefont {Coppersmith},\ and\
  \citenamefont {Eriksson}}]{Goswami:2007p41}%
  \BibitemOpen
  \bibfield  {author} {\bibinfo {author} {\bibfnamefont {S.}~\bibnamefont
  {Goswami}}, \bibinfo {author} {\bibfnamefont {K.~A.}\ \bibnamefont
  {Slinker}}, \bibinfo {author} {\bibfnamefont {M.}~\bibnamefont {Friesen}},
  \bibinfo {author} {\bibfnamefont {L.~M.}\ \bibnamefont {McGuire}}, \bibinfo
  {author} {\bibfnamefont {J.~L.}\ \bibnamefont {Truitt}}, \bibinfo {author}
  {\bibfnamefont {C.}~\bibnamefont {Tahan}}, \bibinfo {author} {\bibfnamefont
  {L.~J.}\ \bibnamefont {Klein}}, \bibinfo {author} {\bibfnamefont {J.~O.}\
  \bibnamefont {Chu}}, \bibinfo {author} {\bibfnamefont {P.~M.}\ \bibnamefont
  {Mooney}}, \bibinfo {author} {\bibfnamefont {D.~W.}\ \bibnamefont {van~der
  Weide}}, \bibinfo {author} {\bibfnamefont {R.}~\bibnamefont {Joynt}},
  \bibinfo {author} {\bibfnamefont {S.~N.}\ \bibnamefont {Coppersmith}}, \ and\
  \bibinfo {author} {\bibfnamefont {M.~A.}\ \bibnamefont {Eriksson}},\ }\href
  {\doibase 10.1038/nphys475} {\bibfield  {journal} {\bibinfo  {journal} {Nat.
  Phys.}\ }\textbf {\bibinfo {volume} {3}},\ \bibinfo {pages} {41} (\bibinfo
  {year} {2007})}\BibitemShut {NoStop}%
\bibitem [{\citenamefont {Borselli}\ \emph {et~al.}(2011)\citenamefont
  {Borselli}, \citenamefont {Ross}, \citenamefont {Kiselev}, \citenamefont
  {Croke}, \citenamefont {Holabird}, \citenamefont {Deelman}, \citenamefont
  {Warren}, \citenamefont {Alvarado-Rodriguez}, \citenamefont {Milosavljevic},
  \citenamefont {Ku}, \citenamefont {Wong}, \citenamefont {Schmitz},
  \citenamefont {Sokolich}, \citenamefont {Gyure},\ and\ \citenamefont
  {Hunter}}]{Borselli:2011p123118}%
  \BibitemOpen
  \bibfield  {author} {\bibinfo {author} {\bibfnamefont {M.~G.}\ \bibnamefont
  {Borselli}}, \bibinfo {author} {\bibfnamefont {R.~S.}\ \bibnamefont {Ross}},
  \bibinfo {author} {\bibfnamefont {A.~A.}\ \bibnamefont {Kiselev}}, \bibinfo
  {author} {\bibfnamefont {E.~T.}\ \bibnamefont {Croke}}, \bibinfo {author}
  {\bibfnamefont {K.~S.}\ \bibnamefont {Holabird}}, \bibinfo {author}
  {\bibfnamefont {P.~W.}\ \bibnamefont {Deelman}}, \bibinfo {author}
  {\bibfnamefont {L.~D.}\ \bibnamefont {Warren}}, \bibinfo {author}
  {\bibfnamefont {I.}~\bibnamefont {Alvarado-Rodriguez}}, \bibinfo {author}
  {\bibfnamefont {I.}~\bibnamefont {Milosavljevic}}, \bibinfo {author}
  {\bibfnamefont {F.~C.}\ \bibnamefont {Ku}}, \bibinfo {author} {\bibfnamefont
  {W.~S.}\ \bibnamefont {Wong}}, \bibinfo {author} {\bibfnamefont {A.~E.}\
  \bibnamefont {Schmitz}}, \bibinfo {author} {\bibfnamefont {M.}~\bibnamefont
  {Sokolich}}, \bibinfo {author} {\bibfnamefont {M.~F.}\ \bibnamefont {Gyure}},
  \ and\ \bibinfo {author} {\bibfnamefont {A.~T.}\ \bibnamefont {Hunter}},\
  }\href@noop {} {\bibfield  {journal} {\bibinfo  {journal} {Appl. Phys.
  Lett.}\ }\textbf {\bibinfo {volume} {98}},\ \bibinfo {pages} {123118}
  (\bibinfo {year} {2011})}\BibitemShut {NoStop}%
\bibitem [{\citenamefont {Lim}\ \emph {et~al.}()\citenamefont {Lim},
  \citenamefont {Zwanenberg}, \citenamefont {Yang},\ and\ \citenamefont
  {Dzurak}}]{Lim:2011preprint}%
  \BibitemOpen
  \bibfield  {author} {\bibinfo {author} {\bibfnamefont {W.~H.}\ \bibnamefont
  {Lim}}, \bibinfo {author} {\bibfnamefont {F.~A.}\ \bibnamefont {Zwanenberg}},
  \bibinfo {author} {\bibfnamefont {C.~H.}\ \bibnamefont {Yang}}, \ and\
  \bibinfo {author} {\bibfnamefont {A.~S.}\ \bibnamefont {Dzurak}},\
  }\href@noop {} {\enquote {\bibinfo {title} {Spin filling of valley-orbit
  states in silicon quantum dot},}\ }\bibinfo {note}
  {ArXiv:1103.2895}\BibitemShut {NoStop}%
\bibitem [{\citenamefont {Shaji}\ \emph {et~al.}(2008)\citenamefont {Shaji},
  \citenamefont {Simmons}, \citenamefont {Thalakulam}, \citenamefont {Klein},
  \citenamefont {Qin}, \citenamefont {Luo}, \citenamefont {Savage},
  \citenamefont {Lagally}, \citenamefont {Rimberg}, \citenamefont {Joynt},
  \citenamefont {Friesen}, \citenamefont {Blick}, \citenamefont {Coppersmith},\
  and\ \citenamefont {Eriksson}}]{Shaji:2008p540}%
  \BibitemOpen
  \bibfield  {author} {\bibinfo {author} {\bibfnamefont {N.}~\bibnamefont
  {Shaji}}, \bibinfo {author} {\bibfnamefont {C.~B.}\ \bibnamefont {Simmons}},
  \bibinfo {author} {\bibfnamefont {M.}~\bibnamefont {Thalakulam}}, \bibinfo
  {author} {\bibfnamefont {L.~J.}\ \bibnamefont {Klein}}, \bibinfo {author}
  {\bibfnamefont {H.}~\bibnamefont {Qin}}, \bibinfo {author} {\bibfnamefont
  {H.}~\bibnamefont {Luo}}, \bibinfo {author} {\bibfnamefont {D.~E.}\
  \bibnamefont {Savage}}, \bibinfo {author} {\bibfnamefont {M.~G.}\
  \bibnamefont {Lagally}}, \bibinfo {author} {\bibfnamefont {A.~J.}\
  \bibnamefont {Rimberg}}, \bibinfo {author} {\bibfnamefont {R.}~\bibnamefont
  {Joynt}}, \bibinfo {author} {\bibfnamefont {M.}~\bibnamefont {Friesen}},
  \bibinfo {author} {\bibfnamefont {R.~H.}\ \bibnamefont {Blick}}, \bibinfo
  {author} {\bibfnamefont {S.~N.}\ \bibnamefont {Coppersmith}}, \ and\ \bibinfo
  {author} {\bibfnamefont {M.~A.}\ \bibnamefont {Eriksson}},\ }\href {\doibase
  doi:10.1038/nphys988} {\bibfield  {journal} {\bibinfo  {journal} {Nat.
  Phys.}\ }\textbf {\bibinfo {volume} {4}},\ \bibinfo {pages} {540} (\bibinfo
  {year} {2008})}\BibitemShut {NoStop}%
\bibitem [{\citenamefont {Liu}\ \emph {et~al.}(2008)\citenamefont {Liu},
  \citenamefont {Fujisawa}, \citenamefont {Ono}, \citenamefont {Inokawa},
  \citenamefont {Fujiwara}, \citenamefont {Takashina},\ and\ \citenamefont
  {Hirayama}}]{Liu:2008p073310}%
  \BibitemOpen
  \bibfield  {author} {\bibinfo {author} {\bibfnamefont {H.~W.}\ \bibnamefont
  {Liu}}, \bibinfo {author} {\bibfnamefont {T.}~\bibnamefont {Fujisawa}},
  \bibinfo {author} {\bibfnamefont {Y.}~\bibnamefont {Ono}}, \bibinfo {author}
  {\bibfnamefont {H.}~\bibnamefont {Inokawa}}, \bibinfo {author} {\bibfnamefont
  {A.}~\bibnamefont {Fujiwara}}, \bibinfo {author} {\bibfnamefont
  {K.}~\bibnamefont {Takashina}}, \ and\ \bibinfo {author} {\bibfnamefont
  {Y.}~\bibnamefont {Hirayama}},\ }\href {\doibase 10.1103/PhysRevB.77.073310}
  {\bibfield  {journal} {\bibinfo  {journal} {Phys. Rev. B}\ }\textbf {\bibinfo
  {volume} {77}},\ \bibinfo {pages} {073310} (\bibinfo {year}
  {2008})}\BibitemShut {NoStop}%
\bibitem [{\citenamefont {Kyriakidis}\ \emph {et~al.}(2002)\citenamefont
  {Kyriakidis}, \citenamefont {Pioro-Ladriere}, \citenamefont {Ciorga},
  \citenamefont {Sachrajda},\ and\ \citenamefont
  {Hawrylak}}]{Kyriakidis:2002p035320}%
  \BibitemOpen
  \bibfield  {author} {\bibinfo {author} {\bibfnamefont {J.}~\bibnamefont
  {Kyriakidis}}, \bibinfo {author} {\bibfnamefont {M.}~\bibnamefont
  {Pioro-Ladriere}}, \bibinfo {author} {\bibfnamefont {M.}~\bibnamefont
  {Ciorga}}, \bibinfo {author} {\bibfnamefont {A.~S.}\ \bibnamefont
  {Sachrajda}}, \ and\ \bibinfo {author} {\bibfnamefont {P.}~\bibnamefont
  {Hawrylak}},\ }\href@noop {} {\bibfield  {journal} {\bibinfo  {journal}
  {Phys. Rev. B}\ }\textbf {\bibinfo {volume} {66}},\ \bibinfo {pages} {035320}
  (\bibinfo {year} {2002})}\BibitemShut {NoStop}%
\bibitem [{\citenamefont {Amasha}\ \emph {et~al.}(2008)\citenamefont {Amasha},
  \citenamefont {MacLean}, \citenamefont {Radu}, \citenamefont {Zumbuehl},
  \citenamefont {Kastner}, \citenamefont {Hanson},\ and\ \citenamefont
  {Gossard}}]{Amasha:2008p2332}%
  \BibitemOpen
  \bibfield  {author} {\bibinfo {author} {\bibfnamefont {S.}~\bibnamefont
  {Amasha}}, \bibinfo {author} {\bibfnamefont {K.}~\bibnamefont {MacLean}},
  \bibinfo {author} {\bibfnamefont {I.~P.}\ \bibnamefont {Radu}}, \bibinfo
  {author} {\bibfnamefont {D.~M.}\ \bibnamefont {Zumbuehl}}, \bibinfo {author}
  {\bibfnamefont {M.~A.}\ \bibnamefont {Kastner}}, \bibinfo {author}
  {\bibfnamefont {M.~P.}\ \bibnamefont {Hanson}}, \ and\ \bibinfo {author}
  {\bibfnamefont {A.~C.}\ \bibnamefont {Gossard}},\ }\href {\doibase
  10.1103/PhysRevLett.100.046803} {\bibfield  {journal} {\bibinfo  {journal}
  {Phys. Rev. Lett.}\ }\textbf {\bibinfo {volume} {100}},\ \bibinfo {pages}
  {046803} (\bibinfo {year} {2008})}\BibitemShut {NoStop}%
\bibitem [{\citenamefont {Thalakulam}\ \emph {et~al.}(2010)\citenamefont
  {Thalakulam}, \citenamefont {Simmons}, \citenamefont {Rosemeyer},
  \citenamefont {Savage}, \citenamefont {Lagally}, \citenamefont {Friesen},
  \citenamefont {Coppersmith},\ and\ \citenamefont
  {Eriksson}}]{Thalakulam:2010p183104}%
  \BibitemOpen
  \bibfield  {author} {\bibinfo {author} {\bibfnamefont {M.}~\bibnamefont
  {Thalakulam}}, \bibinfo {author} {\bibfnamefont {C.~B.}\ \bibnamefont
  {Simmons}}, \bibinfo {author} {\bibfnamefont {B.~M.}\ \bibnamefont
  {Rosemeyer}}, \bibinfo {author} {\bibfnamefont {D.~E.}\ \bibnamefont
  {Savage}}, \bibinfo {author} {\bibfnamefont {M.~G.}\ \bibnamefont {Lagally}},
  \bibinfo {author} {\bibfnamefont {M.}~\bibnamefont {Friesen}}, \bibinfo
  {author} {\bibfnamefont {S.~N.}\ \bibnamefont {Coppersmith}}, \ and\ \bibinfo
  {author} {\bibfnamefont {M.~A.}\ \bibnamefont {Eriksson}},\ }\href@noop {}
  {\bibfield  {journal} {\bibinfo  {journal} {Appl. Phys. Lett.}\ }\textbf
  {\bibinfo {volume} {96}},\ \bibinfo {pages} {183104} (\bibinfo {year}
  {2010})}\BibitemShut {NoStop}%
\bibitem [{\citenamefont {Simmons}\ \emph {et~al.}(2007)\citenamefont
  {Simmons}, \citenamefont {Thalakulam}, \citenamefont {Shaji}, \citenamefont
  {Klein}, \citenamefont {Qin}, \citenamefont {Blick}, \citenamefont {Savage},
  \citenamefont {Lagally}, \citenamefont {Coppersmith},\ and\ \citenamefont
  {Eriksson}}]{Simmons:2007p213103}%
  \BibitemOpen
  \bibfield  {author} {\bibinfo {author} {\bibfnamefont {C.~B.}\ \bibnamefont
  {Simmons}}, \bibinfo {author} {\bibfnamefont {M.}~\bibnamefont {Thalakulam}},
  \bibinfo {author} {\bibfnamefont {N.}~\bibnamefont {Shaji}}, \bibinfo
  {author} {\bibfnamefont {L.~J.}\ \bibnamefont {Klein}}, \bibinfo {author}
  {\bibfnamefont {H.}~\bibnamefont {Qin}}, \bibinfo {author} {\bibfnamefont
  {R.~H.}\ \bibnamefont {Blick}}, \bibinfo {author} {\bibfnamefont {D.~E.}\
  \bibnamefont {Savage}}, \bibinfo {author} {\bibfnamefont {M.~G.}\
  \bibnamefont {Lagally}}, \bibinfo {author} {\bibfnamefont {S.~N.}\
  \bibnamefont {Coppersmith}}, \ and\ \bibinfo {author} {\bibfnamefont {M.~A.}\
  \bibnamefont {Eriksson}},\ }\href {\doibase 10.1063/1.2816331} {\bibfield
  {journal} {\bibinfo  {journal} {Appl. Phys. Lett.}\ }\textbf {\bibinfo
  {volume} {91}},\ \bibinfo {pages} {213103} (\bibinfo {year}
  {2007})}\BibitemShut {NoStop}%
\bibitem [{\citenamefont {Hada}\ and\ \citenamefont
  {Eto}(2003)}]{Hada:2003p155322}%
  \BibitemOpen
  \bibfield  {author} {\bibinfo {author} {\bibfnamefont {Y.}~\bibnamefont
  {Hada}}\ and\ \bibinfo {author} {\bibfnamefont {M.}~\bibnamefont {Eto}},\
  }\href@noop {} {\bibfield  {journal} {\bibinfo  {journal} {Phys. Rev. B}\
  }\textbf {\bibinfo {volume} {68}},\ \bibinfo {pages} {155322} (\bibinfo
  {year} {2003})}\BibitemShut {NoStop}%
\bibitem [{\citenamefont {Elzerman}\ \emph {et~al.}(2004)\citenamefont
  {Elzerman}, \citenamefont {Hanson}, \citenamefont {van Beveren},
  \citenamefont {Vandersypen},\ and\ \citenamefont
  {Kouwenhoven}}]{Elzerman:2004p731}%
  \BibitemOpen
  \bibfield  {author} {\bibinfo {author} {\bibfnamefont {J.~M.}\ \bibnamefont
  {Elzerman}}, \bibinfo {author} {\bibfnamefont {R.}~\bibnamefont {Hanson}},
  \bibinfo {author} {\bibfnamefont {L.~H.~W.}\ \bibnamefont {van Beveren}},
  \bibinfo {author} {\bibfnamefont {L.~M.~K.}\ \bibnamefont {Vandersypen}}, \
  and\ \bibinfo {author} {\bibfnamefont {L.~P.}\ \bibnamefont {Kouwenhoven}},\
  }\href {\doibase 10.1063/1.1757023} {\bibfield  {journal} {\bibinfo
  {journal} {Appl. Phys. Lett.}\ }\textbf {\bibinfo {volume} {84}},\ \bibinfo
  {pages} {4617} (\bibinfo {year} {2004})}\BibitemShut {NoStop}%
\bibitem [{\citenamefont {Thalakulam}\ \emph {et~al.}(2011)\citenamefont
  {Thalakulam}, \citenamefont {Simmons}, \citenamefont {Bael}, \citenamefont
  {Rosemeyer}, \citenamefont {Savage}, \citenamefont {Lagally}, \citenamefont
  {Friesen}, \citenamefont {Coppersmith},\ and\ \citenamefont
  {Eriksson}}]{Thalakulam:2011p045307}%
  \BibitemOpen
  \bibfield  {author} {\bibinfo {author} {\bibfnamefont {M.}~\bibnamefont
  {Thalakulam}}, \bibinfo {author} {\bibfnamefont {C.~B.}\ \bibnamefont
  {Simmons}}, \bibinfo {author} {\bibfnamefont {B.~J.~V.}\ \bibnamefont
  {Bael}}, \bibinfo {author} {\bibfnamefont {B.~M.}\ \bibnamefont {Rosemeyer}},
  \bibinfo {author} {\bibfnamefont {D.~E.}\ \bibnamefont {Savage}}, \bibinfo
  {author} {\bibfnamefont {M.~G.}\ \bibnamefont {Lagally}}, \bibinfo {author}
  {\bibfnamefont {M.}~\bibnamefont {Friesen}}, \bibinfo {author} {\bibfnamefont
  {S.~N.}\ \bibnamefont {Coppersmith}}, \ and\ \bibinfo {author} {\bibfnamefont
  {M.~A.}\ \bibnamefont {Eriksson}},\ }\href@noop {} {\bibfield  {journal}
  {\bibinfo  {journal} {Phys. Rev. B}\ }\textbf {\bibinfo {volume} {84}},\
  \bibinfo {pages} {045307} (\bibinfo {year} {2011})}\BibitemShut {NoStop}%
\bibitem [{\citenamefont {Hanson}\ \emph {et~al.}(2007)\citenamefont {Hanson},
  \citenamefont {Kouwenhoven}, \citenamefont {Petta}, \citenamefont {Tarucha},\
  and\ \citenamefont {Vandersypen}}]{Hanson:2007p1217}%
  \BibitemOpen
  \bibfield  {author} {\bibinfo {author} {\bibfnamefont {R.}~\bibnamefont
  {Hanson}}, \bibinfo {author} {\bibfnamefont {L.~P.}\ \bibnamefont
  {Kouwenhoven}}, \bibinfo {author} {\bibfnamefont {J.~R.}\ \bibnamefont
  {Petta}}, \bibinfo {author} {\bibfnamefont {S.}~\bibnamefont {Tarucha}}, \
  and\ \bibinfo {author} {\bibfnamefont {L.~M.~K.}\ \bibnamefont
  {Vandersypen}},\ }\href {\doibase 10.1103/RevModPhys.79.1217} {\bibfield
  {journal} {\bibinfo  {journal} {Rev. Mod. Phys.}\ }\textbf {\bibinfo {volume}
  {79}},\ \bibinfo {pages} {1217} (\bibinfo {year} {2007})}\BibitemShut
  {NoStop}%
\bibitem [{\citenamefont {Friesen}\ and\ \citenamefont
  {Coppersmith}(2010)}]{Friesen:2010p115324}%
  \BibitemOpen
  \bibfield  {author} {\bibinfo {author} {\bibfnamefont {M.}~\bibnamefont
  {Friesen}}\ and\ \bibinfo {author} {\bibfnamefont {S.~N.}\ \bibnamefont
  {Coppersmith}},\ }\href {\doibase 10.1103/PhysRevB.81.115324} {\bibfield
  {journal} {\bibinfo  {journal} {Phys. Rev. B}\ }\textbf {\bibinfo {volume}
  {81}},\ \bibinfo {pages} {115324} (\bibinfo {year} {2010})}\BibitemShut
  {NoStop}%
\bibitem [{\citenamefont {Culcer}\ \emph {et~al.}(2010)\citenamefont {Culcer},
  \citenamefont {Hu},\ and\ \citenamefont {Das~Sarma}}]{Culcer:2010p205315}%
  \BibitemOpen
  \bibfield  {author} {\bibinfo {author} {\bibfnamefont {D.}~\bibnamefont
  {Culcer}}, \bibinfo {author} {\bibfnamefont {X.}~\bibnamefont {Hu}}, \ and\
  \bibinfo {author} {\bibfnamefont {S.}~\bibnamefont {Das~Sarma}},\ }\href@noop
  {} {\bibfield  {journal} {\bibinfo  {journal} {Phys. Rev. B}\ }\textbf
  {\bibinfo {volume} {82}},\ \bibinfo {pages} {205315} (\bibinfo {year}
  {2010})}\BibitemShut {NoStop}%
\bibitem [{\citenamefont {Boykin}\ \emph {et~al.}(2004)\citenamefont {Boykin},
  \citenamefont {Klimeck}, \citenamefont {Eriksson}, \citenamefont {Friesen},
  \citenamefont {Coppersmith}, \citenamefont {von Allmen}, \citenamefont
  {Oyafuso},\ and\ \citenamefont {Lee}}]{Boykin:2004p115}%
  \BibitemOpen
  \bibfield  {author} {\bibinfo {author} {\bibfnamefont {T.~B.}\ \bibnamefont
  {Boykin}}, \bibinfo {author} {\bibfnamefont {G.}~\bibnamefont {Klimeck}},
  \bibinfo {author} {\bibfnamefont {M.~A.}\ \bibnamefont {Eriksson}}, \bibinfo
  {author} {\bibfnamefont {M.}~\bibnamefont {Friesen}}, \bibinfo {author}
  {\bibfnamefont {S.~N.}\ \bibnamefont {Coppersmith}}, \bibinfo {author}
  {\bibfnamefont {P.}~\bibnamefont {von Allmen}}, \bibinfo {author}
  {\bibfnamefont {F.}~\bibnamefont {Oyafuso}}, \ and\ \bibinfo {author}
  {\bibfnamefont {S.}~\bibnamefont {Lee}},\ }\href {\doibase 10.1063/1.1637718}
  {\bibfield  {journal} {\bibinfo  {journal} {Appl. Phys. Lett.}\ }\textbf
  {\bibinfo {volume} {84}},\ \bibinfo {pages} {115} (\bibinfo {year}
  {2004})}\BibitemShut {NoStop}%
\bibitem [{\citenamefont {Saraiva}\ \emph {et~al.}(2010)\citenamefont
  {Saraiva}, \citenamefont {Koiller},\ and\ \citenamefont
  {Friesen}}]{Saraiva:2010p245314}%
  \BibitemOpen
  \bibfield  {author} {\bibinfo {author} {\bibfnamefont {A.~L.}\ \bibnamefont
  {Saraiva}}, \bibinfo {author} {\bibfnamefont {B.}~\bibnamefont {Koiller}}, \
  and\ \bibinfo {author} {\bibfnamefont {M.}~\bibnamefont {Friesen}},\
  }\href@noop {} {\bibfield  {journal} {\bibinfo  {journal} {Phys. Rev. B}\
  }\textbf {\bibinfo {volume} {82}},\ \bibinfo {pages} {245314} (\bibinfo
  {year} {2010})}\BibitemShut {NoStop}%
\bibitem [{\citenamefont {Gamble}\ \emph {et~al.}()\citenamefont {Gamble},
  \citenamefont {Coppersmith},\ and\ \citenamefont
  {Friesen}}]{Gamble:2011preprint}%
  \BibitemOpen
  \bibfield  {author} {\bibinfo {author} {\bibfnamefont {J.~K.}\ \bibnamefont
  {Gamble}}, \bibinfo {author} {\bibfnamefont {S.~N.}\ \bibnamefont
  {Coppersmith}}, \ and\ \bibinfo {author} {\bibfnamefont {M.}~\bibnamefont
  {Friesen}},\ }\href@noop {} {}\bibinfo {note} {To be submitted.}\BibitemShut
  {Stop}%
\end{thebibliography}


\end{document}